\documentstyle[11pt,epsfig,aaspp4]{article}
\begin{document}

\title{Accretion-Induced Conversion of High-Velocity Neutron Stars 
to Strange Stars in Supernovae and Implications for Gamma-Ray Bursts}

\author{K. S. Cheng$^1$ and Z. G. Dai$^2$} 
\affil{$^1$Department of Physics, University of Hong Kong, Hong Kong, 
China \\
$^2$Department of Astronomy, Nanjing University, Nanjing 210093,
China}

\begin{abstract}
We present a new model for gamma-ray bursts (GRBs) that are not only 
associated with supernovae but also have small baryon contamination.
In this model, we assume a newborn neutron star to move outward 
at a kick velocity of $\sim 10^3\,{\rm km}\,\,{\rm s}^{-1}$ in the supernova 
ejecta. We find that such a neutron star still hypercritically accretes 
its surrounding supernova matter. Once the stellar mass increases to 
some critical mass, the neutron star will undergo a phase transition 
to become a strange star, leading to an energy release of a few 
$10^{52}$ ergs. The phase transition, if possibly occuring just near 
the supernova front, will first result in an ultra-relativistic fireball and 
then a GRB. This provides a plausible explanation for the GRB-supernova
association. We estimate the burst rate to be $\sim 10^{-6}$ per year 
per galaxy. Our model also predicts other possiblities. 
For example, if the resulting fireballs have a Lorentz factor of the order of 
a few, they will produce X-ray GRBs observed by BeppoSAX. We find 
the rate of such bursts to be $\sim 10^{-5}$ per year per galaxy.       
\end{abstract}
 
\keywords{gamma-rays: bursts --- stars: neutron --- supernovae: general}

\section{Introduction}

Gamma-ray bursts (GRBs) emit an amount of isotropic-equivalent energy 
$E_{\rm iso}\ge 10^{52}$ ergs in $\gamma$-rays and X-rays in a few
seconds and subsequently emit afterglows at X-ray (Costa et al. 1997), optical
(van Paradijs et al. 1997) and radio bands (Frail et al. 1997), which generally
last days to months (van Paradijs, Kouveliotou \& Wijers 2000). The energetics
of GRBs, which is comparable to that of supernovae, and their rapid variability 
strongly suggest compact objects involving black holes, neutron stars and 
strange stars as the energy source for GRBs. Two popular models 
satisfying this energetics are explosive events of very massive 
stars, also named hypernovae (Paczy\'nski 1998) or collapsars (Woosley 1993;
MacFadyen \& Woosley 1999), and mergers of neutron-star binaries
(Eichler et al. 1989: Narayan, Paczy\'nski \& Piran 1992). Other possible
models include phase transitions of neutron stars to strange stars 
(Cheng \& Dai 1996; Dai \& Lu 1998a; Bombaci \& Datta 2000; Wang et al. 2000), 
births of magnetars (Usov 1992; Klu\'zniak \& Ruderman 1998; Spruit 1999;
Wheeler et al. 2000), and implosions of supra-massive neutron stars  
(Vietri \& Stella 1998, 1999). 

There has been considerable evidence linking 
the progenitors of GRBs with massive stars.  
First,  the sources of GRBs with known redshifts lie within the optical 
radii and central regions of the host galaxies rather than far outside the disks of 
the galaxies (Bloom, Kulkarni \& Djorgovski 2001), which seem to rule out mergers 
of neutron-star binaries as the GRB central engine. Second,  the brightness 
distribution of GRBs is in agreement with the models in which the GRB rate 
tracks the star formation rate over the past 15 billion years of cosmic 
history (Totani 1997; Wijers et al. 1998; Kommers et al. 2000). Third, the 
supernova SN1998bw with an unusual brightness was discovered in the 
error box of GRB 980425 (Galama et al. 1998) and a supernova-like 
component was detected in the afterglows from GRB 980326 (Bloom et al. 1999) 
and GRB 970228 (Reichart 1999; Galama et al. 2000b). These detections 
provide the most direct evidence for the relation between GRBs and 
a specific type of supernova. Finally, the recent discovery 
of a transient absorption edge in the X-ray spectrum of GRB 990705 
(Amati et al. 2000) and the observations of X-ray lines from GRB 991216 
(Piro et al. 2000) and GRB 000214 (Antonelli et al. 2000) provide new evidence 
that GRBs are related to the core collapse of massive stars. Therefore,
it has been widely believed that long duration GRBs arise from the 
explosions of massive stars. 

However, it seems theoretically difficult to understand the association of 
GRBs with supernovae. This is because the rapid variability of GRBs and their 
nonthermal spectra (Woods \& Loeb 1995) requires that the Lorentz factor of 
a GRB fireball be $\Gamma\ge 100$. This conclusion was also
recently drawn by Lithwick \& Sari (2001), who derived the lower limits 
on $\Gamma$ due to annihilation of photons and scattering of photons 
by pair-created electrons and positrons. Thus, the mass of 
the baryons contaminating the fireball must be less than $M_0 \sim 
10^{-5}M_\odot(\Gamma/500)^{-1}(E_{\rm iso}/10^{52}{\rm ergs})$.
On the other hand,  too many baryons in the stellar envelope exist in 
the vicinity of the collapsing core in a massive star so that 
an ultrarelativistic fireball or jet forming during the collapse of the core 
is easy to become non-relativistic and then a bubble due to 
sideways expansion of the jet. A choked fireball (or jet) has been found 
numerically by MacFadyen \& Woosley (1999) 
     
Here we propose a scenario for the formation of GRBs when a newborn 
neutron star accretes sufficient mass to undergo a  phase transition
to a strange star. Cheng \& Dai (1996), Dai \& Lu (1998a), Bombaci \& Datta 
(2000), and Wang et al. (2000) have proposed the conversion from neutron 
stars to strange stars as a cosmological origin of GRBs, but have not 
investigated the association of GRBs with supernovae, the burst rate, and 
other interesting implications, e.g., X-ray GRBs. The purpose of this paper is 
to discuss these questions. We assume the neutron star to move outward 
at an initial kick velocity of $\sim 10^3\,{\rm km}\,\,{\rm s}^{-1}$ in the supernova 
ejecta. In section 2, we analyze and calculate hypercritically accreted mass. 
In section 3, we discuss implications for GRBs.
In order to produce a GRB, the phase transition is required to occur near 
the supernova front. We estimate the burst rate to be $\sim 10^{-6}$ 
per year per galaxy. Our model explains the GRB-supernova
connection. In section 4, we discuss other implications of the model.
       
\section{Accretion of High-Velocity Neutron Stars in Supernovae}

It is known that the core collapse of massive stars with 
$10-25M_\odot$ produces Type II supernovae accompanying neutron stars 
whose initial mass is likely near the Chrandrasehkar limit $\sim 1.4M_\odot$. 
The neutron stars are believed to have proper velocities. 
Gunn \& Ostriker (1970) first recognized that Galactic pulsars 
have much larger random velocities than their progenitor massive stars. 
Modern observations and analysis on the  proper motion of pulsars 
even give $\sim 450\,{\rm km}\,\,{\rm s}^{-1}$ as an average 
3-dimension velocity of neutron stars at birth (e.g., Lyne \& Lorimer 1994; 
Lorimer et al. 1997; Hansen \& Phinney 1997; Cordes \& Chernoff 1998), 
with possibly a significant population having velocities greater than 
$1000\,{\rm km}\,\,{\rm s}^{-1}$. Direct evidence for pulsar velocities 
$\ge 1000\,{\rm km}\,\,{\rm s}^{-1}$ is provided by observations of 
the bow shock produced by PSR B$2224+65$ in the interstellar medium  
(Cordes, Romani \& Lundgren 1993). The studies of the associations
of neutron stars with supernova remnants have, in many cases, 
indicate large velocities (e.g., Frail et al. 1994). In particular, 
the associations of soft gamma-ray repeaters (SGRs) with supernova 
remnants imply that SGR $0526-66$ and SGR $1900+14$ have velocities 
of $\sim 2900(3\,{\rm kyr}/t_{\rm SNR})\,{\rm km}\,\,{\rm s}^{-1}$ and $\sim 
1800(10\,{\rm kyr}/t_{\rm SNR})\,{\rm km}\,\,{\rm s}^{-1}$ respectively
where $t_{\rm SNR}$ is the supernova remnant age, 
although the associations seem problematic. Since many isolated pulsars 
have such large proper velocities, it appears necessary to invoke 
``natal kicks" imparted to newborn neutron stars due to asymmetrical 
processes during supernovae. Several mechanisms have been 
suggested for natal kicks: local hydrodynamical instabilities, 
neutrino - magnetic field driven asymmetry, local high-order gravity 
mode instabilities, and electromagnetic radiation of an off-centered 
rotating magnetic dipole (for a recent review see Lai 2000). Owing to 
these mechanisms, we assume a newborn neutron star to have an 
initial kick velocity of $v_{\rm ns}\sim 10^3\,{\rm km}\,\,{\rm s}^{-1}$. 

The supernova explosion scenarios involve an outgoing shock wave
that initially travels with speed of $\sim 10^4\,{\rm km}\,\,{\rm s}^{-1}$.
However, when the initial outgoing shock wave enters hydrogen 
envelope, a deceleration of matter is formed (Woosely 1988). This 
deceleration sharpens into a reverse shock. As a result, the final
velocity of the supernova ejecta may be slowed down to $v_{\rm sn}
\sim 10^3\,{\rm km}\,\,{\rm s}^{-1}$. For simplicity, we assume that, 
since this time, the expanding supernova ejecta is spherical, its 
mass $M_{\rm ej}\sim 10M_\odot$ is constant with time, its density $\rho$ is 
uniform in space and decreases with time, and the supernova front radius 
$R$ increases at the fixed velocity of $v_{\rm sn}$. Therefore, the density
throughout is given by
\begin{equation}
\rho=\frac{3M_{\rm ej}}{4\pi R^3}=4.8(R_{0,11}+v_{\rm sn,8}t_3)^{-3}\,{\rm g}
\,\,{\rm cm}^{-3},
\end{equation}
where $v_{\rm sn,8}=v_{\rm sn}/10^8\,{\rm cm}\,\,{\rm s}^{-1}$,
$R_0=R_{0,11}\times 10^{11}$ cm is the initial radius of the ejecta, and 
$t_3$ is the time after the explosion in units of $10^3$ s. Since the ejecta
is radiation dominated, its temperature $T$ scales as $\propto R^{-3/4}$. 
From this scaling law, we have $T=6.5\times 10^7(R_{0,11}+v_{\rm sn,8}t_3)^{-3/4}$\,K
(Brown \& Weingartner 1994), and the sound speed of the ejecta
\begin{equation}
c_s=\left(\frac{5kT}{12m_{\rm H}}\right)^{1/2}=0.47\times 10^8
(R_{0,11}+v_{\rm sn,8}t_3)^{-3/8}\,{\rm cm}\,\,{\rm s}^{-1}.
\end{equation} 

The initial outgoing supernova shock may produce a hole with radius of 
$R_{\rm in}\sim 2\times 10^4$ km (Bethe 1993). So, no matter is
accreted by the neutron star in a time of $t_0\equiv R_{\rm in}/v_{\rm ns}=
20R_{\rm in,*}v_{\rm ns,8}^{-1}$ s where $R_{\rm in,*}=R_{\rm in}
/(2\times 10^4{\rm km})$. Subsequently, the neutron 
star will enters the supernova ejecta. We want to calculate
the accretion rate as follows. If the effect of the neutron star magnetic 
field is neglected (in fact, an initial strong magnetic field can rapidly 
decay due to hypercritical accretion, cf. Geppert, Page \& Zannias 1999), 
the accretion rate of the neutron star at radius $r\equiv v_{\rm ns}t$ is 
given by the Bondi-Hoyle accretion formula,
\begin{equation}
\dot{M}=\frac{\rho R_s^2c^4}{v_{\rm tot}^3}\equiv 
\frac{\rho R_s^2c^4}{\{[v_{\rm ns}-v_{\rm sn}(r,t)]^2+c_s^2\}^{3/2}},
\end{equation}
where $R_s=2GM_{\rm ns}/c^2$ is the Schwarzschild radius of the neutron 
star with $M_{\rm ns}\sim 1.4M_\odot$, and $v_{\rm sn}(r,t)=v_{\rm sn}r
/(R_0+v_{\rm sn}t)=v_{\rm sn}v_{\rm ns}t/(R_0+v_{\rm sn}t)$ is the velocity of the 
supernova matter (not front) at radius $r$ (Chevalier 1989). Here we have defined 
$v_{\rm tot}\equiv \{[v_{\rm ns}-v_{\rm sn}(r,t)]^2+c_s^2\}^{1/2}$. We first estimate
the accreted mass. To do this, $v_{\rm tot}$ is assumed to be approximately 
constant. Thus, at early times $t_0 \ll t\ll 10^3(R_{0,11}/v_{\rm sn,8})\,\,{\rm s}$, 
the accretion rate is approximated by
\begin{equation}
\dot{M}\sim 3.2\times 10^{-4}(M_{\rm ns}/1.4M_\odot)^2R_{0,11}^{-3}v_{\rm tot,8}^{-3}
                \,\,M_\odot\,\,{\rm s}^{-1};
\end{equation}
at late times $t\gg 10^3(R_{0,11}/v_{\rm sn,8})\,\,{\rm s}$, the accretion rate 
becomes  
\begin{equation}   
\dot{M}\sim 3.2\times 10^{-4}(M_{\rm ns}/1.4M_\odot)^2v_{\rm tot,8}^{-3}t_3^{-3}           
                 \,\,M_\odot\,\,{\rm s}^{-1}.
\end{equation}
This accretion rate is at least ten orders of magnitude larger than the 
Eddington accretion rate for a solar-mass star. The gravitational energy
released during such a hypercritical accretion is carried away by neutrinos.
It is neutrino emission that allows accretion of the star at a much higher 
rate than the Eddington rate. From simple analytical arguments, Chevalier 
(1989) and Brown \& Weingartner (1994) estimated a lower limit to steady 
neutron star accretion with neutrino losses assuming spherical symmetry: 
$\dot{M}_{\rm cr}\sim 2\times 10^{-5}M_\odot\,{\rm yr}^{-1}$. Our estimated
accretion rate exceeds $\dot{M}_{\rm cr}$ for $t\le 10^6$ s. We estimate 
the accreted mass
\begin{eqnarray}
\Delta M_{\rm acc} & = & \int^t_{t_0}\dot{M}dt=\int^{t_3}_{t_0/10^3}\frac{0.32M_\odot 
(M_{\rm ns}/1.4M_\odot)^2}{(R_{0,11}+v_{\rm sn,8}t_3)^3v_{\rm tot,8}^3}dt_3\\
& \sim & 0.32M_\odot (M_{\rm ns}/1.4M_\odot)^2v_{\rm tot,8}^{-3}(0.5
+R_{0,11}^{-3}-0.5t_3^{-2}),
\end{eqnarray}
where equation (7) is obtained by substituting equations (4) and (5) into (6).
Next, we numerically calculate equation (6). Figure 1 presents the 
accreted mass as a function of time.  It can be seen from this figure that 
the neutron star will be able to accrete considerable matter (with mass of
$\ge 0.5M_\odot$) before its conversion to a strange star.

\section{Conversion of Accreting Neutron Stars to Strange Stars and 
Gamma-Ray Bursts}

Since conversion of neutron stars to strange stars was suggested as
a possible origin of cosmological GRBs by Cheng \& Dai (1996), resulting
rotating strange stars with strong magnetic fields have been proposed
to explain the observed features of some GRB afterglows by some authors
(Dai \& Lu 1998a, 1998b, 2000, 2001; Zhang \& M\'esz\'aros 2001). 
Bodmer (1971) and Witten (1984) conjectured that strange quark matter
may be the true ground state of hadrons. Detailed calculations based
on the zero-temperature thermodynamics of strange matter show that
strange matter is indeed more stable than $^{56}$Fe for a wide range
of  the parameters of the MIT bag model (Farhi \& Jaffe 1984). If this hypothesis
is true, strange stars as a kind of compact object may exist in the Universe.
How are strange stars produced? One natural way is direct collapse 
of the core of a massive star to a strange star. If this way is possible, 
a binary including a strange star and a more massive compact object 
imaginably exists. During the coalescence of such a binary, the strange 
star is disrupted by its companion. As a result, the entire galaxy can be 
comtaminated and all ``neutron stars" become strange stars (Madsen 1988; 
Caldwell \& Friedman 1991;  Klu\'zniak 1994), which is referred to as the 
Madsen-Caldwell-Friedman (MCF) effect. This effect conflicts with the post-glitch 
behavior of pulsars, which is well described by the neutron-superfluid vortex
creep theory, and current strange star models cannot explain the observed  
pulsar glitches (Alpar 1987; Alpar, Pines \& Cheng 1990). 
Another way to produce strange stars is that
a neutron star in a low-mass X-ray binary accretes sufficient mass to undergo
a phase transition to a strange star (Cheng \& Dai 1996). This way can avoid 
the MCF effect. In this paper, we propose a third way to produce strange 
stars, i.e., a newborn ``unbound" neutron star with a kick
velocity of $\sim 10^3\,{\rm m}\,\,{\rm s}^{-1}$ will catch up with the outgoing 
supernova ejecta and will accrete considerable matter so that the neutron 
star can convert to a strange star (see section 2). If such a rapidly moving 
neutron star arises from the supernova explosion of a massive star in 
a binary, the neutron star must be able to escape from the binary system 
because of its too large kick velocity, and thus the resulting strange star 
will have no companion. 
If a newborn neutron star does not have an enough large 
proper velocity to catch up with the supernova ejecta, the star may accrete 
only a mass of $\sim 0.1M_\odot$ (Chevalier 1989) so that it cannot undergo 
a phase transition to become a strange star. 
Therefore, the third way can also avoid the MCF effect. Cheng \& Dai (1998, 
2001) have argued that such a strange star, if it has a strong magnetic field
and a superconducting core, may produce soft gamma-ray repeaters in an 
age of $\sim 10^4$ yrs. Here we discuss its implications for GRBs.  

The conversion of a neutron star to a strange star requires the formation 
of a strange matter seed, which is produced through the deconfinement
of neutron matter at a density of $(7-9)\rho_0$ (where $\rho_0$ is the
saturation nuclear matter density) (Baym 1991), much larger than the central
density of a $1.4M_\odot$ neutron star with a moderately stiff to stiff 
equation of state (as implied by some astrophysical processes, 
cf. Dai \& Lu 1998a). To reach the deconfinement density, Cheng \& Dai (1996)
suggested, a $1.4M_\odot$ neutron star with a moderately stiff to stiff 
equation of state should accrete matter with mass of $\Delta M_{\rm acc}
\ge 0.5M_\odot$ before its phase transition to a strange star. Once the
accreted mass is $\Delta M_{\rm acc}$, a strange matter seed may appear 
in the core of the neutron star, and subsequently the strange matter will
begin to swallow its surrounding neutron matter in a hydrodynamically unstable
mode (detonation). Thus, the neutron star will convert to a strange star
 in a timescale of the order of 0.1 ms. The phase transition includes two 
processes: (1) the neutron matter converts to two-flavor quark matter, and
(2) the two-flavor quark matter converts to strange (three-flavor) quark matter.
The latter process can release the energy per baryon of a few tens of MeV 
in a timescale of $\sim 10^{-7}$ s (Dai et al. 1995). Owing to this process,
the resulting strange star will be as hot as a few $10^{11}$ K. How is 
the energy release due to the phase transition deposited? One mechanism for 
the energy deposition is the neutrino-antineutrino annihilation process
$\nu+\bar{\nu}\rightarrow e^-+e^+$ and the neutrino absorption processes
$\nu_e+n\rightarrow p+e^-$ and $\bar{\nu}_e+p\rightarrow n+e^+$ 
(in the crust). Another mechanism is the creation of electron/positron
pairs in an extremely strong strong electric field at the quark surface 
(Usov 1998, 2001). The total energy deposition for these two mechanisms, 
$E$, is a few $10^{52}$ ergs, which will inevitably lead to a fireball composed 
of $\gamma$ and electron/positron pairs polluted by a small number 
of baryons. The contaminating baryon mass is up to the crustal mass 
of the strange star, $\sim M_0\sim 10^{-5}M_\odot$. Therefore,
the resulting fireball should be accelerated to an ultra-relativistic phase
with Lorentz factor of $\Gamma\ge E/M_0\sim 500(E/10^{52}{\rm ergs})
(M_0/10^{-5}M_\odot)^{-1}$.               
 
The newborn strange star is surrounded by the supernova ejecta matter.  
If the SN ejecta mass in the solid angle of $\sim 2\pi$ between the star 
and the SN front, $\Delta M_{\rm ej}$, greatly exceeds $M_0$, the resulting 
fireball cannot still be accelerated to ultra-relativistic. An ultra-relativistic fireball
also requires $\Delta M_{\rm ej}\le M_0\sim 10^{-5}M_\odot$. We assume that 
$\Delta R$ is the minimum distance of the initial site of the strange star to 
the supernova front. Assuming $\Delta R\ll R$, we estimate the ratio of 
$\Delta M_{\rm ej}$ to the total SN ejecta mass $M_{\rm ej}$: 
\begin{equation}
\frac{\Delta M_{\rm ej}}{M_{\rm ej}}\sim \frac{\pi (\Delta R)^2R\rho}{\pi R^3\rho}
=\left(\frac{\Delta R}{R}\right)^2.
\end{equation} 
Since $\Delta M_{\rm ej}/M_{\rm ej}\le 10^{-6}$, we have 
\begin{equation}
\frac{\Delta R}{R}\le 10^{-3},
\end{equation}
which further requires 
\begin{equation}
\delta_v\equiv \frac{v_{\rm ns}-v_{\rm sn}}{v_{\rm sn}}\le 10^{-3}.
\end{equation}
Because the number of the neutron stars with $v_{\rm ns}\ge 10^8\,{\rm cm}\,\,
{\rm s}^{-1}$ is $N_{\rm ns}(v_{\rm ns}\ge 10^8\,{\rm cm}\,\,{\rm s}^{-1})\sim 10^7$ 
per galaxy in the Hubble time (Lorimer et al. 1997), the number of the strange 
stars that can produce GRBs are estimated to be $N_{\rm ss\rightarrow GRB}
\sim \delta_v\times N_{\rm ns}(v_{\rm ns}\ge 10^8\,{\rm cm}\,\,{\rm s}^{-1})
\sim 10^4$ per galaxy in the Hubble time, where we have assumed that 
the number distribution of the neutron stars is uniform in the velocity space 
for $v_{\rm ns}\ge 10^8\,{\rm cm}\,\,{\rm s}^{-1}$. The burst rate in our model is 
approximated by
\begin{equation}
{\cal{R}}\sim \frac{N_{\rm ss\rightarrow GRB}}{t_{\rm Hubble}}\sim 10^{-6}
\,\,{\rm yr}^{-1}\,\,{\rm per\,\,galaxy}.
\end{equation}
This rate is enough to explain the observed GRB rate, $10^{-7}$ /yr/galaxy. 
The latter rate has been estimated due to the evidence that the GRB
rate is proportional to the star formation rate (Totani 1997; Wijers et al. 1998;
Kommers et al. 2000). 

Since the accreted mass $\Delta M_{\rm acc}\le 1.4M_\odot$, accretion of 
the pre-conversion neutron star should not  influence 
its velocity significantly, we have an approximate relation:
$\Delta R\sim R_0(v_{\rm sn}/v_{\rm ns})$, and the ejecta radius
$R\ge 10^3R_0(v_{\rm sn}/v_{\rm ns})$. Thus, the expansion timescale of the 
SN ejecta is
\begin{equation}
t\sim R/v_{\rm sn}\ge 10^6R_{0,11}v_{\rm ns,8}^{-1}\,\,{\rm s}.
\end{equation}
This implies that the SN explosion might occur a few days before a GRB.

\section{Discussions}

Our model has several implications. First, it can clearly explain the 
association of GRBs with a special kind of supernova. The phase transition 
of the neutron star to a strange star is almost isotropic; about one half of 
the energy release will result in an ultra-relativistic fireball with baryon 
contamination of $\Delta M_{\rm ej}+M_0$, and another half will be absorbed 
by the SN ejecta because the photon-electron scattering depth 
in the SN ejecta $\tau\sim \sigma_T[M_{\rm ej}/(m_p\pi R^3)]R\sim 10^5 \gg 1$. 
So, the phase transition discussed here may also give rise to a hypernova 
with a high amount of explosive energy ($\sim 10^{52}$ ergs) and  
a bright optical luminosity. A hypernova-like component has been 
observed in several cases, e.g., SN 1998bw (Iwamoto et al. 1998), GRB 980326 
(Bloom et al. 1999) and GRB 970228 (Reichart 1999; Galama et al. 2000). 
Second, the observed breaks in the light curves of some optical afterglows 
are argued to be due to sideways expansion of collimated fireballs 
(Rhoads 1999; Sari, Piran \& Halpern 1999). However, this argument  
is still now a matter of considerable theoretical debate (Moderski,
Sikora \& Bulik 2000; Wei \& Lu 2000). Dai \& Lu (1999, 2000) have proposed 
the evolution of a fireball to the non-relativistic regime in a dense medium 
as an alternative explanation for these observed breaks. Our present 
model meets the second explanation. Third, if the phase transition occurs 
at $10^{-3}\ll \Delta R/R\sim 10^{-2}$, then the resulting fireball has 
the loading baryon mass of $\sim (\Delta R/R)^2M_{\rm ej}\sim 10^{-3}M_\odot
(10^2\Delta R/R)^2(M_{\rm ej}/10M_\odot)$, and its Lorentz factor should be
$\sim 5(E/10^{52}\,{\rm ergs})(10^2\Delta R/R)^{-2}(M_{\rm ej}/10M_\odot)^{-1}$. 
Although this fireball cannot produce a classical GRB due to its low Lorentz
factor, it may result in a weak GRB like GRB 980425 or an X-ray GRB 
(or called an X-ray flash) like GRB 991106 observed by BeppoSAX 
(search http://www.ias.rm.cnr.it/ias-home/sax/xraygrb.html).
We predict that the rate of such a kind of GRB is $\sim 10^{-5}(10^2\Delta R/R)$
per year per galaxy. Finally, the phase transition occurs more possibly 
at $\Delta R/R\gg 10^{-2}$. All the energy release will be absorbed by 
the SN ejecta, and the resulting fireball is non-relativistic but much 
more energetic than a normal supernova. This fireball may only behave 
as a hypernova but not emit a GRB or even an X-ray GRB.

%Several solutions to the baryon contamination problem have been 
%suggested, e.g., two-step explosions (Cheng \& Dai 2001a), neutrino 
%oscillations (Klu\'zniak 1998), and supranovae (Vietri \& Stella 1998). 
%In the supranova model, a GRB occurs when a supra-massive neutron star
%loses so much angular momentum through magnetic dipole radiation that 
%the star implodes to a black hole (Vietri \& Stella 1998). Recent observations 
%on the X-ray line in GRBs support such models (Amati et al. 2000; Piro et al.
%2000; Antonelli 2001). We note that the supranova model assumes rapidly 
%rotating supramassive neutron stars. One natural way to produce such 
%stars is accretion, e.g., the Eddington accretion in low-mass X-ray binaries 
%(Vietri \& Stella 1999) or the hypercritical accretion in supernovae (Chevalier 1989). 
%A direct result of mass accretion is the decay of the neutron star magnetic 
%fields by at least 3 to 4 orders of magnitude (Phinney \& Kulkarni 1994; 
%Geppert et al. 1999). Thus, the spin-down time for the stars 
%becomes $t_{\rm sd}\sim 10^7(B/10^9{\rm G})^{-2}\,$yr where $B$ is the stellar
%magnetic field after sufficient accretion. This timescale is clearly too long to 
%account for the association of GRBs with supernovae. 

In summary, we have presented a new model for GRBs with small baryon 
contamination. A key point of our model is that a newborn neutron star with 
an initial kick velocity of $\sim 10^3\,{\rm km}\,\,{\rm s}^{-1}$ will catch up with
the outgoing supernova ejecta, and accrete matter at a hypercritical rate. 
Once the stellar mass increases to some critical mass, the neutron star 
will undergo a phase transition to a strange star, resulting in an energy 
release of a few $10^{52}$ ergs. The phase transition may produce 
a GRB when it occurs just near the SN front. The burst rate is 
$\sim 10^{-6}$ per year per galaxy. In addition, if the phase transition
occurs in the interior of the SN ejecta, it may result in an X-ray GRB 
observed by BeppoSAX or only a burstless hypernova explosion.   

In this paper, we have discussed one result for accreting high-velocity 
neutron stars in supernovae, i.e., the conversion to strange stars 
as an origin of GRBs.  Another result for accreting neutron stars
is the collapse to black holes when the stellar mass reaches the maximum 
mass. If the pre-collapse neutron stars  are millisecond pulsars, the
resulting black holes must be rapidly rotating. As suggested by Vietri \& Stella
(1998), such black holes may produce GRBs by extracting their rotational 
energy or tapping the binding energy of the disk-black hole system.

\acknowledgments
This work was supported by a RGC grant of Hong Kong government, 
the National Natural Science Foundation of China (grant 19825109), and
the National 973 Project (NKBRSF G19990754).

\clearpage
\begin{figure}
\begin{picture}(100,250)
\put(0,0){\includegraphics{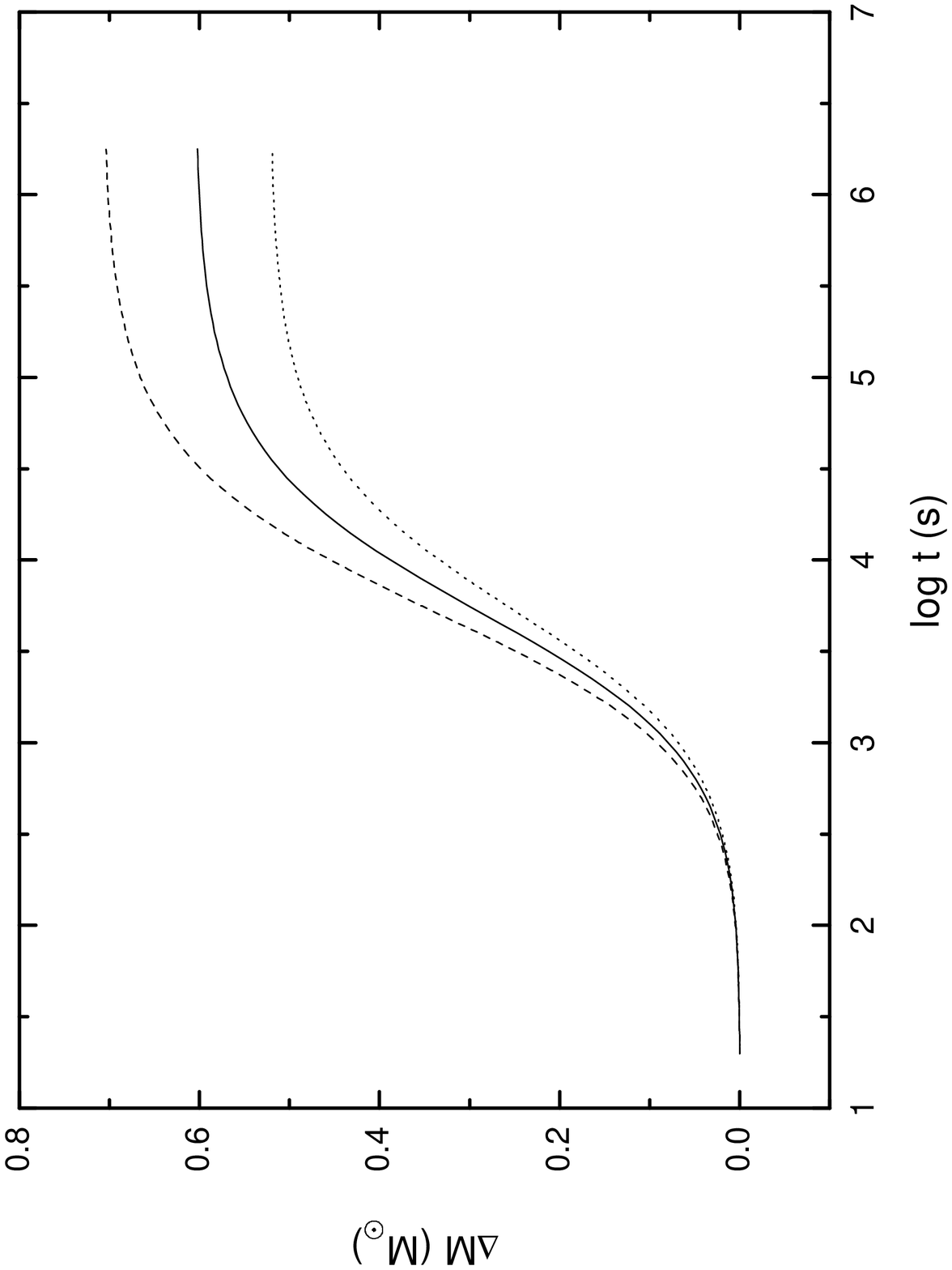}}
\end{picture}
\caption
{The accreted mass as a function of time for $R_{0,11}=1$ and
$M_{\rm ej}=10M_\odot$. The dashed, solid and dotted lines correspond
to $v_{\rm ns,8}=v_{\rm sn,8}=1.2$, 1.3 and 1.4 respectively.}
\end{figure}
 
\end{document}